# RapidBrachyMCTPS 2.0: A Comprehensive and Flexible Monte Carlo-Based Treatment Planning System for Brachytherapy Applications.

Harry Glickman[1], Majd Antaki[1], Christopher Deufel[2], Shirin A. Enger[1,3,4,5]
[1]Medical Physics Unit, McGill University, Montreal, Canada
[2]Department of Radiation Oncology, Mayo Clinic, Rochester, MN 55905
[3]Department of Oncology, McGill University, Montreal, Canada
[4]Research Institute of the McGill University Health Centre, Montreal, Canada
[5]Lady Davis Institute for Medical Research, Jewish General Hospital, Montreal, Canada

**Abstract:** We have previously described RapidBrachyMCTPS, a brachytherapy treatment planning toolkit consisting of a graphical user interface (GUI) and a Geant4-based Monte Carlo (MC) dose calculation engine. This work describes the tools that have recently been added to RapidBrachyMCTPS, such that it now serves as the first stand-alone application for MC-based brachytherapy treatment planning. Notable changes include updated applicator import and positioning, three-plane contouring tools, and updated dose optimization algorithms that, in addition to optimizing dwell position and dwell time, also optimize the rotating shield angles in intensity modulated brachytherapy. The main modules of RapidBrachyMCTPS were validated including DICOM import, applicator import and positioning, contouring, material assignment, source specification, catheter reconstruction, EGSphant generation, interface with the MC code, and dose optimization and analysis tools. Two patient cases were simulated to demonstrate these principles, illustrating the control and flexibility offered by RapidBrachyMCTPS for all steps of the treatment planning pathway. RapidBrachyMCTPS is now a stand-alone application for brachytherapy treatment planning, and offers a user-friendly interface to access powerful MC calculations. It can be used to validate dose distributions from clinical treatment planning systems or model-based dose calculation algorithms, and is also well suited to testing novel combinations of radiation sources and applicators, especially those shielded with high-Z materials.

## I. Introduction

In the past decade, brachytherapy treatments have evolved from surgical source placements with dosimetry based on lookup tables to image-guided remote afterloading source placements using complex, computer-based dose-calculation algorithms and dose optimization techniques. Current dosimetric approaches are based on the American Association of Physicists in Medicine (AAPM) Task Group No. 43 (TG-43) formalism (1), which describes dose deposition around a single source centrally positioned in a spherical water phantom with unit density. Although the TG-43 formalism improved the dosimetry of photon emitting brachytherapy sources over prior dose calculation formalisms, it relies on source-specific data pre-calculated in a standard homogeneous water geometry and disregards patient-specific scatter conditions and radiological differences between different tissue, applicator and source materials from that of water. Indeed, ignoring patient tissue and applicator heterogeneities with TG-43 based dose calculations can yield substantial discrepancies between prescribed and delivered dose for low and high energy brachytherapy sources (2, 3, 4, 5, 6, 7).

The updated guidelines for dosimetry in brachytherapy by AAPM TG-186 (8) recommend the use of model-based dosimetry calculations algorithms (MBDCAs). MBDCAs such as the Monte Carlo (MC) method, collapsed-cone, superposition/convolution, and deterministic solutions to the linear Boltzmann transport equation use information from patient's images for dose calculations. These algorithms provide a detailed and accurate method for calculation of absorbed dose in heterogeneous systems such as the human body, with the MC method being the most accurate and the gold standard. Several commercial treatment planning systems (TPS) have incorporated MBDCAs such as Acuros™ (BrachyVision™ (BV); Varian® Medical Systems, Palo Alto, CA, 9, 10, 11) and Oncentra-ACE™ (Oncentra™ Brachy v. 4.4, Elekta Brachytherapy, Veenendaal, Netherlands, 12, 13) to enable transition from TG-43 to TG-186 allowing dose calculations in heterogeneous media.

Recent technological innovations in brachytherapy such as intensity modulated brachytherapy as well as investigation of novel brachytherapy radiation sources, source models and applicators require researcher input and modification of the commercial MBDCA. In both Acuros and Oncentra-ACE TPSs, however, all calculation settings are pre-set. Radiation source and applicator modification or introduction of novel radiation sources and applicators by the user is not possible. Numerous MC dose calculation engines are published, including PTRAN (14), MCNP (15), MCPI (16), BrachyDose (17), ALGEBRA (18), HDRMC, (19), and egs_brachy (20). However, the majority of these software packages are not publicly available. In addition, these software packages lack contouring tools or optimization algorithms and thus do not represent truly stand-alone MC-based TPSs. Furthermore, many of these engines have hard-coded photon emission spectra for commonly used brachytherapy radiation sources, making it difficult to investigate new radionuclides for brachytherapy applications.

We have developed the first MC-based TPS for brachytherapy applications that can be used as an investigational open platform to push forward innovations in brachytherapy, or used as a second check to TG-186 based calculations with commercial MBDCAs. We have previously described RapidBrachyMCTPS, and validated its Geant4-based MC dose calculation engine (21). The aim of the current work was to describe the contouring tools, catheter reconstruction and optimization algorithms that have been added to RapidBrachyMCTPS, such that it now serves as a stand-alone application for MC-based brachytherapy treatment planning.

## II. Methods

### II.I. Treatment Planning Workflow

The standard of care in brachytherapy is to acquire a 3D image set, such as computed tomography (CT) or magnetic resonance images, and generate a plan that considers the patient's anatomy on the day of treatment (22). This workflow requires catheter positions to be outlined, target and organs at risk (OARs) contoured, dose from each potential dwell position calculated, and the optimal combination of dwell positions and dwell times determined. MC-based treatment planning additionally requires that the material compositions and mass densities of the target, OARs, radiation source, applicators, or any other foreign object inserted in the patient's body to be specified. RapidBrachyMCTPS offers modules facilitating every step of this treatment planning process, providing the users without programming knowledge access to powerful MC-based treatment planning and dosimetry. We describe below the tools implemented in RapidBrachyMCTPS which enable this treatment planning workflow. Once complete, the clinical plan, catheter reconstruction, and structure sets can be exported in DICOM format. Resulting RT Struct files and RT Plan files are compatible with both Oncentra and Eclipse TPSs.

### II.I.I. Patient Image and Density Calibration Curve Import and Export

RapidBrachyMCPTS currently supports importing DICOM image files, RT Struct files, and RT Plan files, including those generated by Elekta or Varian treatment planning systems. The user selects the import directory and all DICOM files are identified and imported. A calibration curve to convert the CT Hounsfield unit (HU) values into mass density must also be provided, listing material/density pairs and associated HU values for CT or synthetic CT. An example calibration curve is provided in Appendix E, Table E2. RT Struct files and RT Plan files can be exported, both compatible with Oncentra and Eclipse TPSs.

### II.I.II. Applicator Import and Positioning

Applicators can be uploaded as a series of .stl files, with each file representing the three dimensional mesh of a continuous single-material applicator component. The components can be translated or rotated together to position them within the CT geometry with click-and-drag functionality. An applicator material definition file is also required with material names, elemental compositions and associated mass densities. Furthermore, an applicator set meta-file can be imported, listing dwell positions, and material assignments for each applicator component separately. Alternatively, users can experiment with applicator material assignments via drop-down menus in the GUI, and choose a custom density for any material. Users can also contour applicators directly with the contouring tools described in section II.I.III.

### II.I.III. Contouring Tools

RapidBrachyMCTPS provides three main tools to create and modify closed polygonal contours: the pen, brush, and morph tools. Branched structures can be represented by multiple closed polygons on the same slice. The pen tool allows users to hold down the left mouse button, adding polygon points along their cursor path. Points can also be placed individually with intermittent clicking, and the polygon can be closed by right clicking. These polygons can be added to or subtracted from contoured structures. Next, the brush tool allows users to hold down the left mouse button, adding polygonal circles to the contoured structure. The radius of the circle is controlled by the user. The brush can also be used as an eraser to subtract from the structure. Finally, the morph tool allows the user to click-and-drag to distort the contoured outline. All points falling within an area-of-effect radius are also pulled along with the clicked coordinate, with their displacement proportional to their proximity. The user has control over this area-of-effect radius. The contour can also be translated without distortion by right-clicking on the contour outline instead of left-clicking. Finally, a paired parametric interpolation tool (see Appendix C) allows users to interpolate all slices along an axis with a single click, or they can interpolate a single slice on a single plane.

Clinical TPSs often only allow for contouring on the axial slices. Patient anatomy is sometimes easier to visualize, however, on the coronal or sagittal slices. Clinicians are therefore encouraged to check their contours on the sagittal, coronal and 3D views to ensure that what they are outlining on the axial slices results in realistic profiles on the other axes. This practice is not always followed. Clinicians sometimes quickly contour on the axial slices and neglect to check the other views, resulting in highly jagged and unrealistic segmentations. This may explain part of the extreme inter-observer variability in clinical contour delineation, which leads to clinically significant errors in radiation planning (23, 24). Our goal was to help reduce these inaccuracies by allowing for contouring on all three orthogonal axes, such that jaggedness can easily be touched up, and clinicians more encouraged to consider all three axes. In RapidBrachyMCTPS, users can utilise any of the tools listed above on any of the three plane view widgets. The details of implementation are found in Appendix C.

The tools described above are polygon-based. MC calculations, however, require labelmap input, a list of which image voxels do or do not belong to a given structure. It is important for the user to understand what discrepancies exist between these two representations, especially for narrow convoluted structures where small differences in voxel assignment can result in large dose differences (25). Pinter *et al* (26) describe the concept of multiple representations of the same structure and the conversions between them. In short, contours are represented in three main ways: planar contours, labelmaps, and 3D surfaces (Figure 1). In RapidBrachyMCTPS, the planar polygon representation is the principle or "master" representation, from which the others are derived. This representation is continually converted to the labelmap representation after every user input, and is immediately displayed. Then, a 3D representation of the contoured structure is generated from this labelmap. Either of these secondary displays can be toggled off.

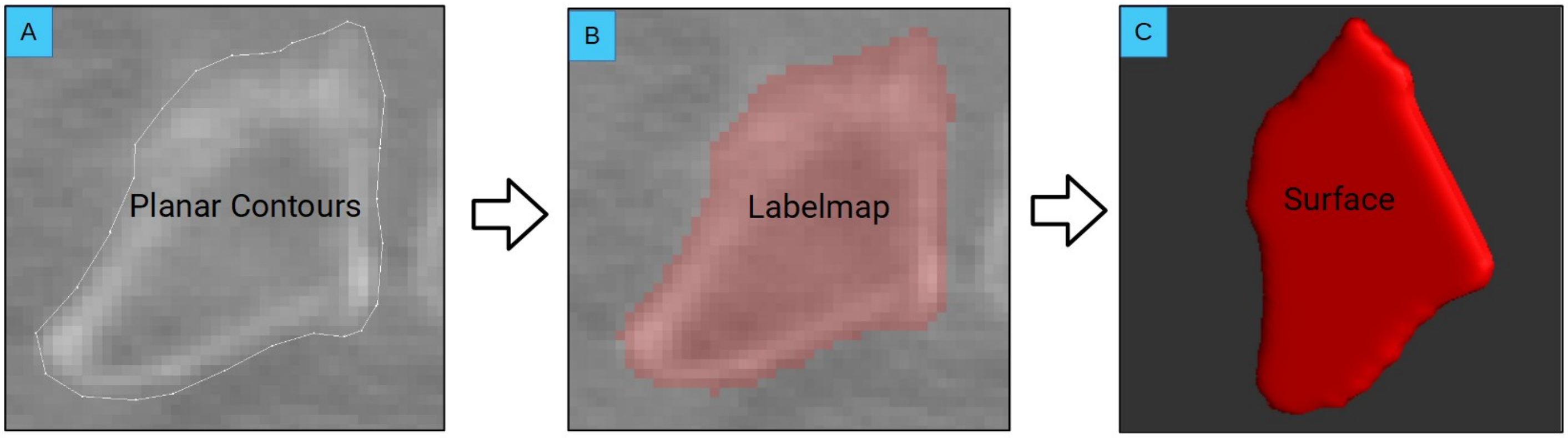

Figure 1: Contour representations and conversions. Planar contours convert to labelmaps, which convert to surfaces. Planar contours are the "master" representation from which other representations are derived. A) planar representation B) labelmap representation C) surface representation.

### II.I.IV. Material and Density Assignment

Materials and mass densities are assigned to contoured or uploaded structures via drop-down menus. For each contour, the user can assign material compositions and mass densities in several ways: A) by assigning material elemental composition and nominal mass density to an entire contoured organ, B) by assigning material elemental composition to the entire contoured organ but assigning voxel-by-voxel mass densities obtained by using a calibration curve, C) by assigning material elemental composition and mass densities voxel-by-voxel using a calibration curve. For MRI-based images, a synthetic CT is created where material composition and nominal mass densities are assigned to each contoured organ. Tissue elemental compositions are implemented according to the TG-186 (8) recommendations.

### II.I.V. Catheter Reconstruction

Manual catheter reconstruction occurs by selecting points one-by-one and adding them to the catheter outline. Dwell positions are sampled at regular intervals from these reconstructed catheters, determined by the user-assigned step size and starting with a user-defined tip offset. Dwell positions can be uploaded as part of a pre-existing RT Plan file, or uploaded with the applicator set meta-file as discussed above. Automatic catheter reconstruction techniques are currently under development and will be added to the TPS after testing.

### II.I.VI. Source Selection

Various commercial source models are implemented in the MC dose calculation engine, RapidBrachyMC. Source geometry is selected from RapidBrachyMC's library via a drop-down menu. Radioactive core material and air kerma rate are also assigned. Any elemental isotope can be selected as the core material, since the radioactive decay for brachytherapy sources is handled through explicit simulation of nuclear decay. Source parameters can be assigned per plan, per catheter, or per dwell.

### II.I.VII. MC Dose Calculation Engine

Patient geometry is imported to RapidBrachyMC in the EGSphant format (27). The original EGSphant format allows only 9 tissue/material types to be assigned to the patient geometry. We have modified the EGSphant format to allow up to 35 tissue/material types. The user selects the voxel dimensions for dose calculations with the MC method, and image data and structure set masks are resampled accordingly. Voxel-by-voxel dose distribution maps are obtainable separately for each dwell position, or a single dose map can be calculated for the entire treatment plan. Several dose scorers are implemented. Dose to medium in medium, dose to water in medium, or dose to water in water can be obtained. The implementation and validation of the MC engine is described in detail by Famulari *et al* (21).

#### II.I.VIII. Optimization

RapidBrachyMCTPS supports forward and inverse optimization. In forward planning, dwell times are changed manually or isodoses are dragged into place. In inverse planning, dwell times are optimized using either mixed integer or column generation optimization techniques. The TPS is equipped with extensive analysis tools for dosimetry described below, facilitating an iterative optimization process. The implementation and validation of the optimizers is described further in Antaki *et al* (28).

#### II.I.IX. Analysis Tools

Isodoses or dose heatmaps can be superimposed over patient images. DVHs are calculated for contoured and imported structures. Dose to medium in medium, dose to water in medium and dose to water in water can all be displayed. As dwell times are updated, whether manually or through optimization, previous DVHs are saved. These DVHs can then be reloaded in the future for comparison. For a given structure, four DVH parameters are displayed: volume percentage receiving at least a given dose, absolute volume receiving at least a given dose, percentage of isodose received by a given volume, and absolute dose received by a given volume. Changing any of these fields updates the remaining fields.

### II.II. Verification

#### II.II.I. DICOM Import

A clinical prostate cancer case comprising patient images in DICOM format, contours in a DICOM RT Structure Set file and the treatment plan in a DICOM RT Plan file were exported from Oncentra TPS and imported into RapidBrachyMCTPS to verify the DICOM import features. To ensure the DICOM world coordinates were interpreted correctly by RapidBrachyMCTPS, relative orientation and position of the patient images, structure set, and catheters were compared between the two software packages. Additionally, the volumes of the structures in the two software packages were compared, as well as with those calculated by SlicerRT (29).

#### II.II.II. MC Simulations

Several simulations were performed with the RapidBrachyMC MC dose calculation engine to validate the capabilities of the GUI. The common details for the MC simulations are described in this section and are summarized in Table E1 according to the recommendations of TG-268 (30). The simulations were run with $10^7$ radioactive decays per dwell position, resulting in type A dose uncertainties below 1% at the 100% isodose line. Radioactive decay was simulated explicitly using the G4RadioactiveDecay class, based on the Evaluated Nuclear Structure Data File (31). Electromagnetic interactions during particle transport were performed by the G4EmPenelopePhysics class, based on the standard Penelope physics list (32). Photon production was cut at 0.1 mm, and electrons were not explicitly transported, as collision kerma in voxels was scored using a track length estimator due to the photon energies emitted from the simulated sources.

##### II.II.II.I. Basic Simulations with $^{192}$Ir and $^{169}$Yb

A solid water phantom (30x30x30 cm$^3$) scanned at McGill University Health Center was used to verify the basic simulation tools: applicator import, applicator positioning, applicator material and density assignment, source core material and density assignment, contour creation, and contour material and density assignment. First, we verified applicator import and positioning. A single-material grooved applicator STL was created in Blender (33) and translated and rotated within the water phantom. The applicator design mesh consists of 642 triangles and can be found in Appendix D. Two initial simulations were performed, one with the applicator material assigned to silicone with mass density of 1.14 g/cm$^3$, and another with the applicator material assigned to tungsten with mass density of 18.1 g/cm$^3$. Material and mass densities of the EGSphant file were assigned voxel-by-voxel from a CT to density calibration curve. A microSelectron v2 source geometry was chosen, with $^{192}$Ir assigned as the active core material. Next, we verified source selection. With the applicator in the same position as the first tests, we assigned the applicator material to tungsten, but the source active core material was changed from $^{192}$Ir to $^{169}$Yb. Finally, a simple cylindrical contour was positioned above the applicator emission window and assigned a material of cortical bone with nominal density assignment of 3.0 g/cm$^3$.

**II.II.II.II. Low Dose Rate Brachytherapy with $^{125}$I**

A hypothetical retinoblastoma treated with low dose rate (LDR) brachytherapy was used to demonstrate the flexibility of RapidBrachyMCTPS, allowing for easy experimentation of novel source material and applicators. We received a 3D computer model of an 18mm diameter eye plaque applicator frame from Trachsel Dental Studio (Rochester, Minnesota). The silastic insert mesh, where brachytherapy seeds are placed, was not provided, therefore we modelled a core to fill the frame in Blender. To demonstrate that novel seed configurations can easily be experimented with, we generated a 9 seed configuration that was different from existing COMS plaques. The applicator design, including the eye plaque mesh, silastic insert mesh, and seed placements, can be found in Appendix D. Since McGill University Health Center does not perform eye plaque brachytherapy, CT images from a head-and-neck cancer patient were used. Globe, retina, optic nerve, and lens were contoured as OARs, and a hypothetical tumour was contoured resting on the retina. The applicator was positioned around the optic nerve. Gold with mass density 19.32 g/cm$^3$ was assigned to the applicator plaque to approximate the Modulay gold alloy and silicone with mass density of 1.14 g/cm$^3$ was assigned to the silastic insert. Patient tissue elemental composition and mass densities were based on a CT to density calibration curve. A SelectSeed (Elekta Brachytherapy, Veenendaal, The Netherlands) source model was used, with the radioactive core material assigned to $^{125}$I.

**II.II.II.III. High Dose Rate Brachytherapy with $^{192}$Ir**

A clinical prostate cancer case was used to demonstrate how a clinical plan can be loaded into RapidBrachyMCTPS, simulated with the MC dose calculation engine with the clinical dwell times (post-implant dosimetry), and then re-optimized with the optimization algorithms implemented in RapidBrachyMCTPS. Catheter positions and structure sets were imported from Oncentra TPS (Elekta Brachytherapy, Veenendaal, The Netherlands). Patient tissue composition and mass densities were assigned voxel-by-voxel from a CT to density calibration curve. The active core material was set to $^{192}$Ir with a microSelectron v2 source geometry. First, the clinical plan dosimetry was calculated using an MC simulation. Next, the plan was re-optimized so as to meet clinical target constraints while sparing OARs, notably creating a low-dose tunnel in the urethra. This re-optimized plan was compared to the clinical plan.

**II.III. Dependencies**

The TPS GUI elements were built using Qt 5.12 (34, 35) a cross-platform UI software development kit. All classes containing GUI elements inherit directly or indirectly from QObject, the main Qt classes as shown in Figure 2. For instance, all the RapidBrachyMCTPS menus inherit from the Menu class, which inherits from QObject. See Appendix A for a full list of the classes comprising RapidBrachyMCTPS.

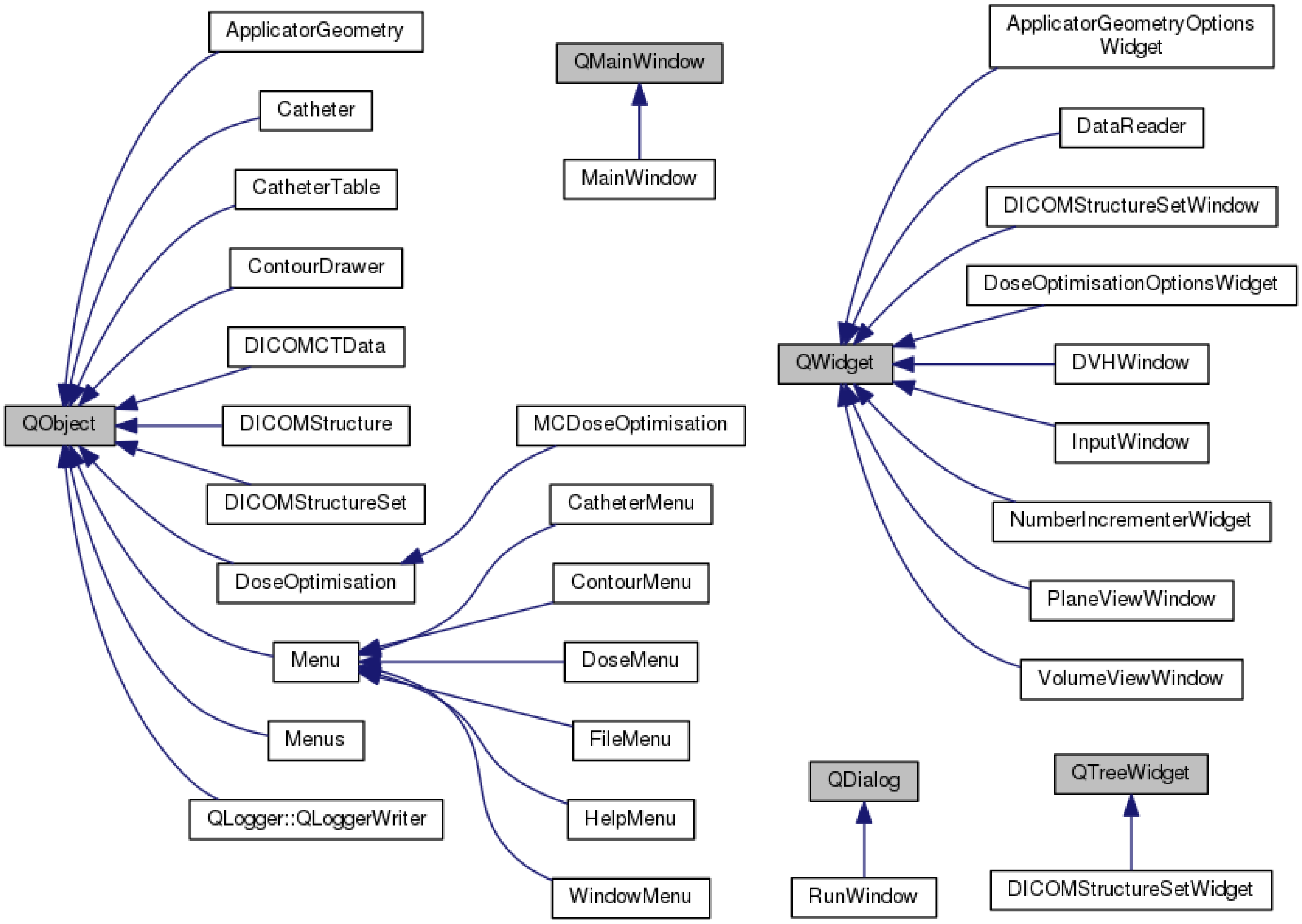

Figure 2: RapidBrachyMCTPS classes inheriting from Qt classes. Qt classes are shaded in grey, RapidBrachyMCTPS classes in white.

Visualization and patient image manipulation were implemented in Vtk 8.2.0 (36, 37) a C++ library designed to manipulate and display scientific data. Vtk is integrated into the Qt GUI using QVTKOpenGLNativeWidget as presented in Figure 3. The plane view widgets and 3d view widgets therefore inherit from this class.

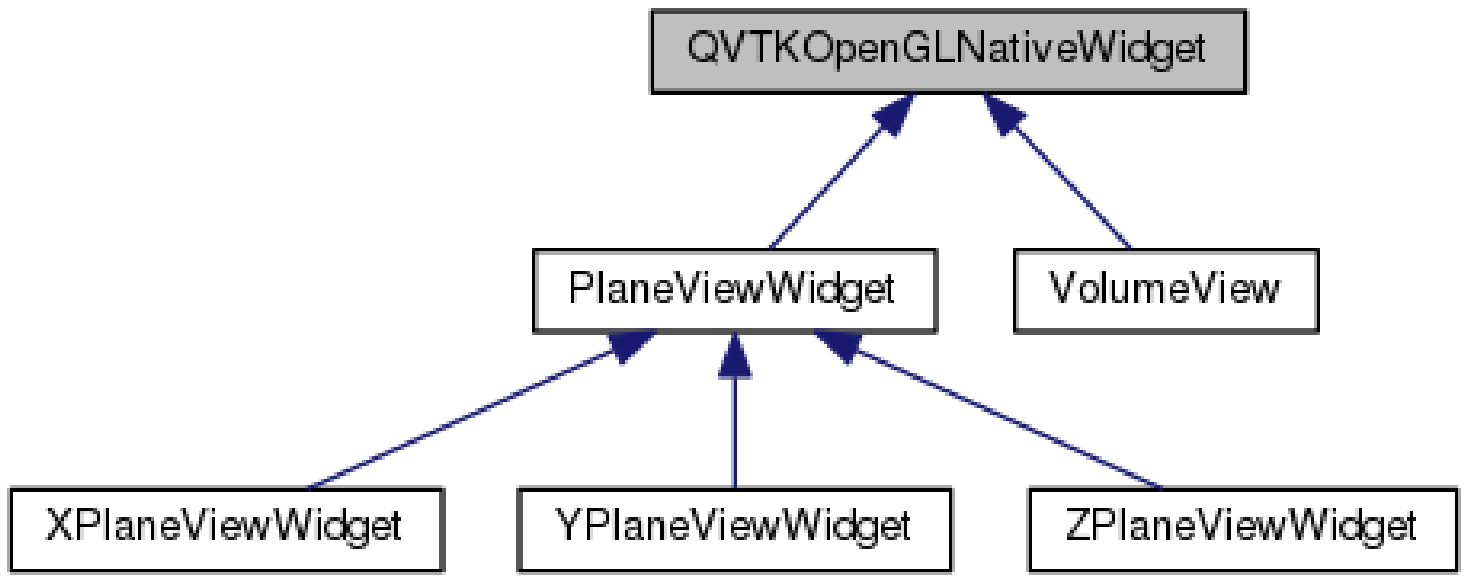

Figure 3: QVTKOpenGLNativeWidget inheritance. QVTKOpenGLNativeWidget allows for incorporation of Vtk-based medical images in Qt-based GUIs, and therefore all image viewing widgets inherit from this class.

Optimization is handled by Gurobi, a commercial quadratic mixed integer optimizer (38). Our implementation applies penalty-based and mixed-integer methods, coupled with techniques to reduce the problem size. Given a set of dosimetric constraints, a single optimal plan is calculated that maximizes dose to target while minimizing dose to organs at risk. Interaction with Gurobi takes place through the MCDoseOptimisation class. For a more detailed discussion of the implementation, see Antaki *et al* (28). Gurobi is freely available for academic purposes, but non-academic users of RapidBrachyMCTPS will need to purchase a commercial license.

DICOM interactions are handled by GDCM (39) described in detail in Appendix B.

## III. Results

Validation of RapidBrachyMCTPS's main software modules are presented below, demonstrating its flexible functionality.

### III.I. DICOM Import

The patient files imported into the RapidBrachyMCTPS main view window are presented in Figure 4A. The three planes views are visible, as is the list of dwell positions. The same files opened in the Oncentra TPS window are presented in Figure 4B. Again, the plane views and dwell positions are visible. Relative positions and orientations of patient image files, structure sets and catheters are preserved throughout the export and import process.

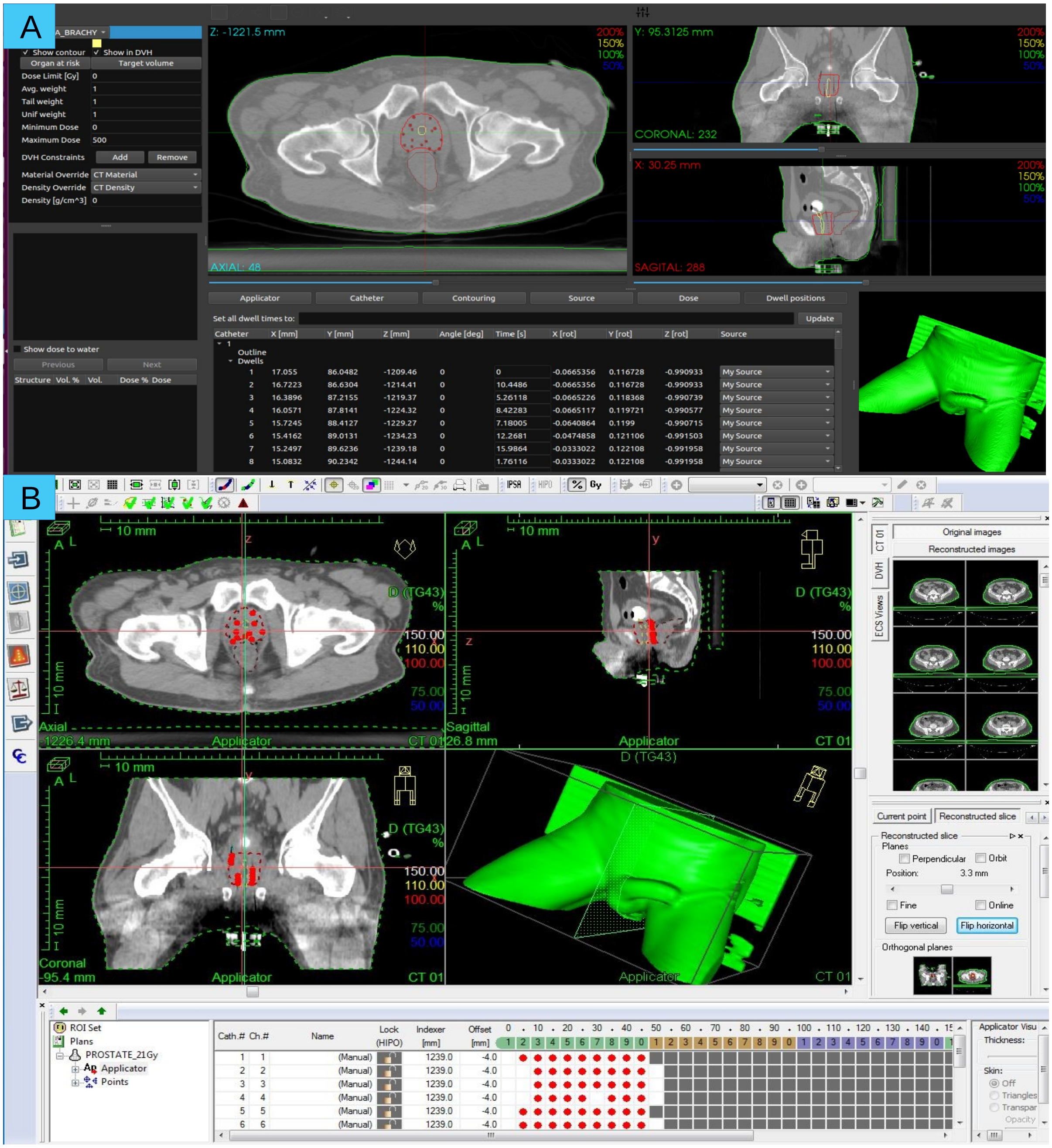

Figure 4: DICOM files exported from Oncentra and imported to RapidBrachyMCTPS. A) RapidBrachyMCTPS Dwell Positions window. B) Oncentra Planning window.

The structure set volumes for RapidBrachyMCTPS, Oncentra and SlicerRT are presented in Table 1. Volumes are nearly identical between RapidBrachyMCTPS and SlicerRT, but vary up to 8% from those calculated by Oncentra.

| **Volumes (cm^3)** *(Calculation method)* | **RapidBrachyMCTPS** *(Voxel-based)* | **Oncentra** *(Distance-map)* | **SlicerRT** *(Voxel-based)* |
|---|---|---|---|
| **Body** | 16788.02 | 16606.35 | 16785.7 |
| **CTV** | 52.46 | 50.79 | 52.4615 |
| **Rectum** | 27.42 | 26.41 | 27.4197 |
| **Urethra** | 2.52 | 2.33 | 2.52 |

Table 1: Structure set volumes form RapidBrachyMCTPS, Oncentra, and SlicerRT.

**III.II. Basic Simulation**

Figure 5 illustrates the dosimetrical results of changing source radioactive core material, applicator material, and contour material: one of these parameters was changed at a time with the other two fixed. First, in Figure 5A, the applicator material was set to silicone, the radioactive core material was set to $^{192}$Ir, and no contour was included, and thus the material assignments to each voxel were determined using the CT density alone. Next, in Figure 5B, the applicator material was changed to tungsten, again with $^{192}$Ir core material and no contour. It can readily be seen that a tungsten applicator has a greater shielding effect than a silicone applicator. Next, in Figure 5C, the source radioactive core material was changed to $^{169}$Yb, with the applicator material again set to tungsten and no contour included. Dose distributions are similar, although from the sagittal profile we observe a narrower beam for $^{169}$Yb as compared to $^{192}$Ir, demonstrating the greater effect of high-Z materials on $^{169}$Yb as compared to $^{192}$Ir. Finally, in Figure 5D, a cylindrical contour was created and material was assigned to cortical bone, with the applicator material set to tungsten and the active core material set to $^{169}$Yb. It can be observed that the bone contour modulated the dose distribution, absorbing a significant amount of dose. Therefore, changing any of the three parameters had the expected result on the dose distribution.

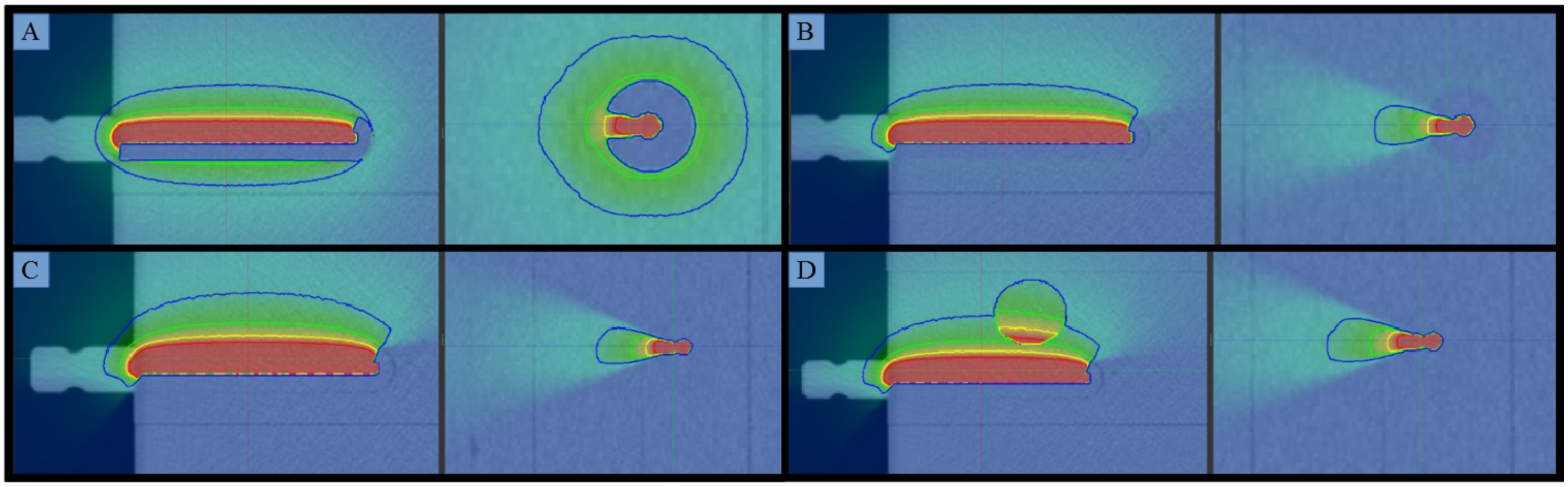

Figure 5: MC simulation results with different applicator, source and contour assignments. Source geometry is microSelectron v2. Dwell times manually scaled to deliver visually comparable dose profile for $^{169}$Yb as compared to $^{192}$Ir. A) Silicone applicator with $^{192}$Ir assigned to the active core, no contour. B) Tungsten applicator with $^{192}$Ir assigned to the active core, no contour. C) Tungsten applicator with $^{169}$Yb assigned to the active core, no contour. D) Tungsten applicator with $^{169}$Yb assigned to the active core, bone assigned to the contour. For all simulations, dose inside the applicator was removed.

**III.III. End-to-End Treatment Planning LDR**

Figure 6 demonstrates the dosimetric results for a hypothetical retinoblastoma case. The applicator geometry was correctly positioned in the patient geometry for the MC simulation. It can be observed that the applicator provides substantial shielding effects, and that the dose is well concentrated in the vicinity of the tumour. No optimization was performed, and therefore the dosimetry could be greatly improved. Note also that the orbit bones received substantial dose. Saving times for all drawn contours were less than 0.5 seconds.

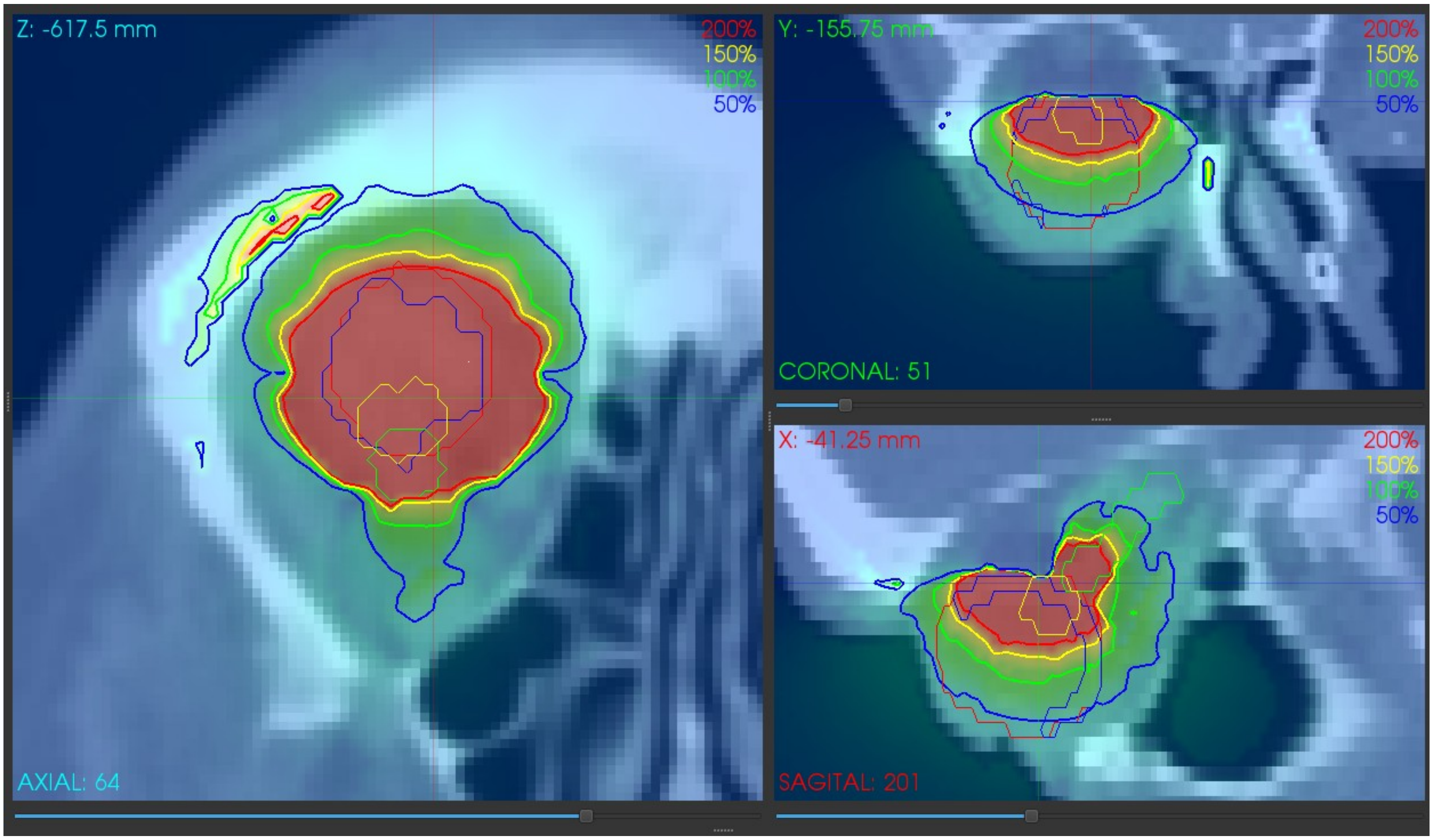

Figure 6: Isodoses and dose colorwash for eye plaque applicator. Source goemetry is SelectSeed with iodine assigned as a radioactive core. Dwell times were manually scaled to deliver approximately a 50 Gy prescription dose around the tumour. The patient was positioned obliquely within the CT scanner, and thus the three orthogonal planes are not the traditional sagittal, coronal, and axial views.

**III.IV. End-to-End Treatment Planning HDR**

Figure 7 demonstrates dosimetric results for the clinical plan re-simulated in RapidBrachyMCTPS. Figure 7A presents the results from the post-implant dosimetry simulation, where the dwell times are taken from the clinical plan. Figure 7B presents the simulation when the dwell times are optimized with the inverse optimization implemented in RapidBrachyMCTPS. This optimization was constrained to create a low-dose tunnel in the urethra, at the expense of tumour coverage. DVH parameters for the two plans are presented in Table 2. Dose to the target remained within clinically acceptable parameters, while OARs were relatively spared.

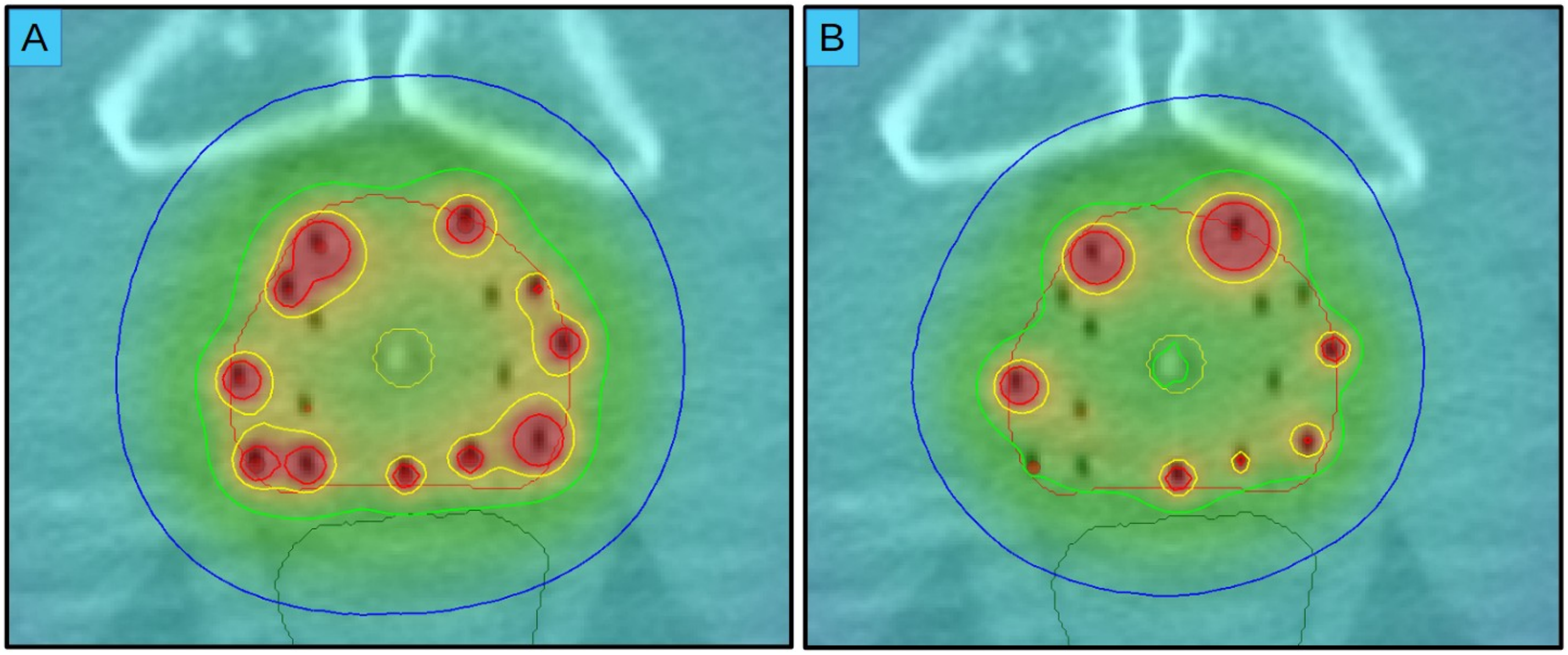

Figure 7: Resulting isodoses and dose colorwash for A) clinical dwell times compared to B) the RapidBrachyMCTPS-optimized dwell times where a low-dose tunnel was created in the urethra.

| **Dosimetric Indices** | **Clinical (Gy)** | **RapidBrachyMCTPS (Gy)** |
|---|---|---|
| Rectum 2cc | 10.28 | 9.16 |
| Bladder 2cc | 10.30 | 9.97 |
| Urethra + margins 0.1cc | 19.65 | 16.23 |
| Urethra 0.1cc | 18.48 | 15.64 |
| Tumour D90 | 17.49 | 15.31 |
| Tumour V100 (%) | 98.69 | 93.82 |

Table 2: Selected DVH parameters from clinical and RapidBrachyMCTPS plans.

## IV. Discussion

The Monte Carlo engine on which RapidBrachyMCTPS is based, RapidBrachyMC, has been robustly validated in a previous study (21). The validations presented above therefore served to demonstrate that the new additions to the developed GUI layer were interacting with this engine appropriately. We also showcased the clinical utility of RapidBrachyMCTPS for both LDR and HDR brachytherapy. All tools worked as expected, with the flexibility to handle a variety of use-cases. The software can be used as an investigational platform to push forward innovations in brachytherapy, or as a way to robustly verify the plans generated by commercial MBDCAs. The user can, for example, investigate any radioisotope as a new source for brachytherapy applications, develop and experiment with novel applicators, or simply benchmark clinical treatment plans. With an open-source release, researchers and clinicians either use the software as-is, or they can add more modules to customize and improve upon the design.

### IV.I. Validations

In the first validation, a clinical prostate cancer case comprising patient CT images, a structure set, and a treatment plan file was exported from Oncentra TPS and imported into RapidBrachyMCTPS. This confirmed that RapidBrachyMCTPS is capable of correctly importing DICOM-format images, DICOM RT Structure Set files, and DICOM RT Plan files. We confirmed that relative positions and orientations for the patient images, structure set, and catheters were correct. As an additional test, we verified that structure set volumes were preserved between the two software packages. Volume differences of up to 8% were observed between RapidBrachyMCTPS and Oncentra, explained by the use of different volume calculation algorithms.

RapidBrachyMCTPS simply calculates volume voxel-wise, considering a voxel inside a structure if the center of that voxel falls within the contour outline. Oncentra TPS, however, uses a distance-map to generate a triangulated surface (Oncentra Brachy Physics and Algorithms User Manual). Volumes calculated in RapidBrachyMCTPS were nearly identical to those generated by SlicerRT, which also uses a voxel-based volume calculation. A more detailed discussion of volume calculation algorithms and their impact on DVH parameters can be found in Kirisits *et al* (40).

In the second validation, we demonstrated that the basic tools enabling simulation were working correctly: applicator positioning, applicator material and density assignment, radioactive core selection, contour material and density assignment, and interface with the MC calculation engine. Applicator material assignment, radioactive core selection, and presence of a bone contour were individually varied. The results demonstrated that higher-Z materials shield photons emitted from $^{192}$Ir and $^{169}$Yb more efficiently than lower-Z materials, and that $^{169}$Yb is more affected by these shielding effects than $^{192}$Ir. Furthermore, high-Z bone tissue absorbed substantial dose. We therefore confirmed that assignment parameters within the GUI had the expected effects on dose distributions.

### IV.II. Showcases

In the first showcase, a hypothetical retinoblastoma case was planned with a $^{125}$I LDR SelectSeed source, demonstrating RapidBrachyMCTPS's flexibility with respect to novel applicators and radiation sources. An applicator mesh was modelled, dwell positions were assigned, a theoretical tumour as well as OARs were contoured, and a treatment plan was generated. The 12772 triangles comprising the eye plaque applicator mesh did not cause any appreciable slowdown on import or applicator positioning. This same pipeline could be used for any applicator geometry, any materials, and any source specification, enabling flexible experimentation with applicator and source combinations.

In the second showcase, we demonstrated a typical use case for prostate HDR brachytherapy. We selected a clinical prostate cancer case from our institution to be re-simulated and re-optimized by RapidBrachyMCTPS. Optimization criteria were set so as to create a low-dose tunnel in the urethra, at the expense of tumour coverage. While planning target volume D90 and V100 in the re-optimized plan remained within clinically acceptable parameters, these indices were lower than those calculated using the clinical dwell times. Differences between prescribed and delivered dose as calculated by MBDCAs and commercial treatment planning systems are well documented, and the magnitude of the difference depends on the photon energy and composition of the involved tissues (41, 7). However, in this study, the clinical plan was re-simulated in medium and the dosimetric differences are due to the re-optimization of the plan. A full discussion regarding the impact of optimization, choices of different dosimetric calculation methods and patient tissue segmentation schedules on dosimetry is beyond the scope of this showcase, and have been explored elsewhere (28, 2, 7).

### IV.III. Limitations and Planned Improvements

While RapidBrachyMCTPS supports all steps of the brachytherapy treatment planning process, its capabilities continue to grow and numerous features are planned for future development. Notably, our lab is developing a deep convolutional neural network algorithm designed to obtain the desired radiation quantities with a high speed and at accuracies arbitrarily close to those of the source MC algorithm (42). We also plan on incorporating automatic catheter reconstruction and planning tools. Smaller planned improvements include cross-platform installation, DICOM RT Dose file import and export, DICOM anonymization, contouring performance improvements, and image set rotation.

### IV.IV. Download and Installation

The RapidBrachyMCTPS source code will be provided open-source, along with a script for installing the software and all requirements on Linux operating systems. These requirements include Root (43), Geant4, Vtk, GDCM, Gurobi, and RapidBrachyMC. Users will need to acquire a Gurobi license, which may be free or paid depending on the entity requesting the license. All other software requirements are free and open-source. Should this script fail to install the software, a separate installation guide is provided, detailing all installation steps. Users can also contact the RapidBrachyMCTPS team to be provided with further installation support.

## V. Conclusion

With the tools described and validated in this work, RapidBrachyMCTPS now serves as the first stand-alone MC-based TPS for brachytherapy applications. The software will be available open-source to researchers and clinicians worldwide. We have demonstrated its power and flexibility for both experimentation and benchmarking. The software is particularly well-suited to testing novel radiation sources and applicators, especially those shielded with high-Z materials.

## Acknowledgements

This work was supported by Collaborative Health Research Projects (grant numbers 523394-18, 248490) and the Natural Sciences and Engineering Research Council of Canada (NSERC) (grant number 241018). Computations were performed on the Calcul Quebec MP2 and Beluga clusters (44), a regional organization of Compute Canada (45).

## Appendix A: RapidBrachyMCTPS classes

The MainWindow class inherits from QMainWindow and is instantiated on startup. It is responsible for initializing many of the other classes, as well as setting up the Qt connections between them. On startup the MainWindow instantiates the classes described in Table A1. These classes in turn instantiate or inherit from the remaining RapidBrachyMCTPS classes, listed in Table A2.

| Class | Function |
|---|---|
| InputWindow | DICOM directory & material table selection |
| Menus (CatheterMenu, ContourMenu, DoseMenu, FileMenu, HelpMenu, WindowMenu) | Menu UIs |
| DICOMCTData | Parses and stores CT Data |
| DICOMStructureSet, DICOMStructureSetWidget, DICOMStructureSetWindow | Maintains, displays list of structure segmentations (each segmentation is a DICOMStructure class) |
| CatheterTable | Maintains list of catheters, dwell positions (each catheter is a Catheter class) |
| ApplicatorGeometry, ApplicatorGeometryOptionsWidget | Maintains list of applicator components |
| MCDoseOptimization, DoseOptimisationOptionsWidget | Enables MC dose calculation |
| ContourDrawer | Enables structure segmentation |
| VolumeView, VolumeViewWindow | Three dimensional structure visualization |
| XplaneViewWidget, YplaneViewWidget, ZplaneViewWidget | Three plane views, each with a PlaneViewWindow |
| DVHWindow | Displays the DVH graph and table |

Table A1: Classes instantiated by MainWindow on startup and their purposes.

| **Class** | **Function** |
|---|---|
| PlaneViewWidget | Parent class of the three PlaneViewWidget classes. |
| DoseOptimisation | Parent class of the MCDoseOptimization class. |
| RunWindow | Interface with RapidBrachyMC |
| NumberIncrementerWidget | Custom implementation of QDoubleSpinBox class. |
| InteractorStyle | Custom vtk interactor, inherits from vtkInteractorStyleImage |
| CustomContourWidget | Custom contouring widget, inherits from vtkContourWidget |
| Qlogger, written by: (46) | Registers and prints log files |

Table A2: Other RapidBrachyMCTPS classes.

## Appendix B: DICOM Compatibility

### B.I. DICOM Format

Medical information, especially medical images, is often transmitted in the DICOM format (47). The format consists of series of DICOM data elements organized hierarchically (Figure B1). These elements contain information about a patient, called the data values, uniquely identified by their DICOM tags. Each tag is represented by two hexadecimal numbers, a group number and an element number, in the format (XXXX, XXXX). Each element also stores a value representation, such as Integer String or Code String, that defines what type of information it contains. Finally, they also store a value length, specifying the byte length of the data element. When a tag holds nested tags, its value representation is Sequence. Groups of tags are conceptually organized into “Modules”, although this organization is not reflected anywhere in the data hierarchy. Some data elements contain unique identifiers (UIDs), meant to uniquely identify DICOM files around the world.

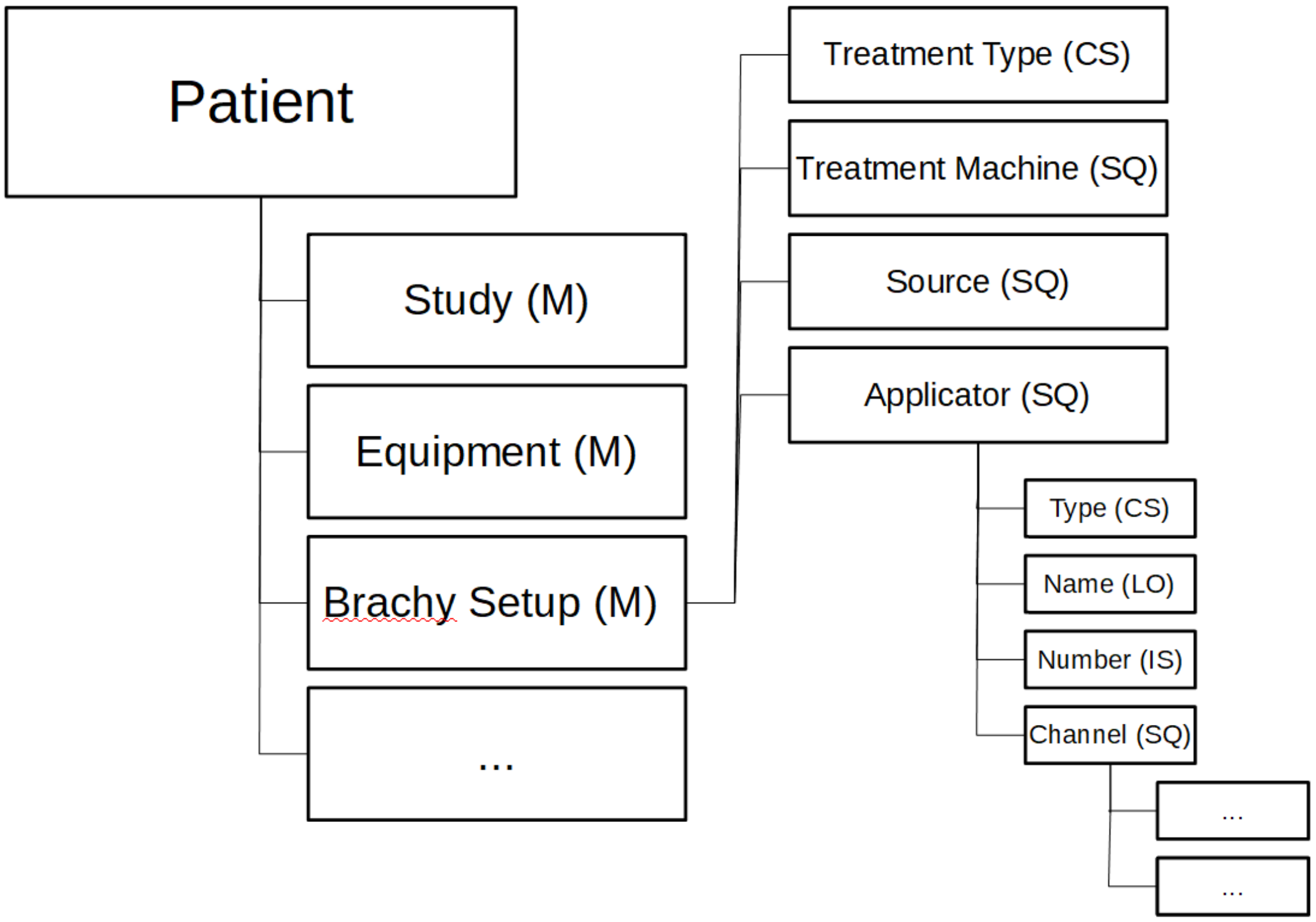

Figure B1: DICOM Specification, tags are organized hierarchically. Tags are conceptually organized into Modules (M), although this is not reflected in the data. Sequence (SQ) tags hold nested tags. Other value representations provide patient information, such as Integer Strings (IS) or Code Strings (CS).

**B.II. Structure Set Generation**

DICOM RT Struct files are created using vtkGDCMPolyDataWriter, with many of the tags copied from the image files. This method is used to generate both the contoured structures and the catheter reconstruction information.

**B.III. Plan Generation**

RT Plan files are created using GDCM's Writer class with all tags, values, value representations (VRs), and value lengths (VLs) set explicitly. Again, many of the tags are copied from the patient image files.

The following modules are mandatory for a brachytherapy RT Plan file:

1) Patient
2) General Study
3) RT Series
4) General Equipment
5) RT General Plan
6) SOP Common
7) RT Brachy Application Setups

**B.III.I. New UIDs created**

Four new UIDs are created:

A) RT Plan UID

Plan Header: 0x0002,0x0003 – Media Storage SOP Instance UID
SOP Common: 0x0008,0x0018 - SOP Instance UID

B) RT Series UID

General Study: 0x0020,0x000e - Series Instance UID

C) RT Study UID

RT Series: 0x0020,0x000d – Study Instance UID

D) Catheter Reconstruction UID

RT General Plan: Referenced Structure Set Sequence: 0x0008,0x1155 – Referenced SOP Instance UID

Struct header: 0x0008,0x0018 - SOP Instance UID

Struct header: 0x0002,0x0003 - SOP Instance UID

They are created by appending to a DICOM Prefix. The current date is appended (year, month, date), followed by the current time (hours, minutes, seconds), followed by a period, followed by a randomly generated number.

**B.III.II. Copied Tags**

The image files are parsed for tags that are common between the image and plan DICOM specifications. The content of these tags is copied and placed in a new DataSet to be used for plan writing. Only the required tags are copied over, optional tags are not (Table B1).

| **Tags** | **Description** |
|---|---|
| **Patient** | |
| 0010,0010 | Patient Name |
| 0010,0020 | Patient ID |
| 0010,0030 | Patient Birth Date |
| 0010,0040 | Patient Sex |
| **General Study** | |
| 0008,0020 | Study Date |
| 0008,0030 | Study Time |
| 0008,0050 | Accession Number |
| 0008,0090 | Referring Physician Name |
| 0020,0010 | Study ID |
| **General Equipment** | |
| 0008,0070 | Manufacturer |

Table B1: Tags copied from image file main dataset.

The header is also copied from the image file, and the following tags are modified: SOP Class UID and SOP Instance UID (Table B2).

| **Tags** | **Description** | **Value** |
|---|---|---|
| 0008,0016 | SOP Common -> SOP Class UID | "1.2.840.10008.5.1.4.1.1.481.5" = plan class |
| 0002,0002 | Header -> Media Storage SOP Class UID | 0x0008,0x0016 |
| 0008,0018 | SOP Common -> SOP Instance UID | Plan UID – see above |
| 0002,0003 | Header -> Media Storage SOP Instance UID | 0x0008,0x0018 |

Table B2: UID tags.

**B.III.III. RT Brachy Applications Setup**

The core of the RT Plan is the RT Brachy Applications Setup module (Table B3). It consists of a description of the source, treatment machine, and applicator. An applicator, source, and treatment machine library are therefore created with information about the following tags. They are written to Treatment Machine Sequence (300A,0206), Source Sequence (300a,0210), and Application Setup Sequence (300a,0230).

| **Tags** | **Description** | **Example Value** |
|---|---|---|
| **Applicator Library** | | |
| 300a, 0200 | Brachy Treatment Technique | "INTRACAVITARY" |
| 300a, 0202 | Brachy Treatment Type | "HDR" |
| **Applicator Library** | **Application Setup Sequence (300a,0230)** | |
| 300a, 0232 | Application Setup Type | "ENDORECTAL" |
| 300a, 0234 | Application Setup Number | 1 |
| 300a, 0250 | Total Reference Air Kerma | “7628.29327534935" |
| *300a,0280* | *Channel Sequence* | See below |
| **Treatment Machine Library** | **Treatment Machine Sequence (300A,0206)** | |
| 300a, 00b2 | Treatment Machine Name | "MUHC HDR v2" |
| 0008, 1090 | Manufacturer Model Name | microSelectron-HDR v2" |
| **Source Library** | **Source Sequence (300a,0210)** | |
| 300a,0212 | Source Number | 0 |
| 300a,0214 | Source Type | "CYLINDER" |
| 300a,0226 | Source Isotope | "Ir-192" |
| 300a,0228 | Source Isotope Half Life | "73.830” |
| 300a,022A | Reference air kerma | "43720.00” |
| 300a,022C | Source Strength Reference Date | "20000101" |
| 300a,022E | Source Strength Reference Time | "000000" |
| 300a, 0218 | Active Source Diameter | “2” |
| 300a, 021a | Active Source Length | “5” |

Table B3: RT Brachy Applications Setup tags.

Each channel has a tag group (Channel Sequence (300a,0280), Table B4), under the Application Setup Sequence (300a,0230).

| **Tags** | **Description** | **Example Value** |
|---|---|---|
| 300a, 0110 | Number Of Control Points | Number of dwells * 2 (to introduce duplication for afterloader) |
| 300a, 0282 | Channel Number | Channel index starting at 1 |
| 300a, 0288 | Source Movement Type | “STEPWISE” |
| 300a, 0290 | Source Applicator Number | 1 |
| 300a, 0291 | Source Applicator ID | 1 |
| 3006, 0084 | Referenced ROI Number | Channel index starting at 0 |
| 300a, 0292 | Source Applicator Type | “FLEXIBLE” |
| 300a, 0296 | Source Applicator Length | “1250” |
| 300a, 02a0 | Source Applicator Step Size | “2.5” |
| 300c, 000e | Referenced Source Number | 1 |
| 300a, 0284 | Channel Length | “1250” |
| 300a, 02c8 | Final Cumulative Time Weight | ** |
| 300a, 0286 | Total Time | ** |
| 300a, 02d0 | Brachy Control Point Sequence | See below |

Table B4: Channel Sequence tags.
**The weights field works as follows: treatment time is (300a, 02d6) times (300a, 0286) divided by (300a, 02c8).

Each control point has a tag group (Brachy Control Point Sequence [300a,02d0], Table B5), under the Channel Sequence [300a,0280].

| **Tags** | **Description** | **Example** |
|---|---|---|
| 300a, 0112 | Control Point Index | Control point index starting at 0 * 2 + duplication index |
| 300a, 02d2 | Control Point Relative Position | Relative position from catheter start |
| 300a, 02d4 | Control Point 3D Position | 3d world coordinates |
| 300a, 02d6 | Cumulative Time Weight in ms | ** |
| 300a, 0412 | Control Point Orientation | Line orientation connecting previous and next control points. *NB: RapidBrachyMCPS writes DS instead of FL Value Representation, not clear why this was necessary.* |

Table B5: Brachy Control Point Sequence tags.

### B.III.IV. Other Tags

Next, there are some additional tags not contained in the image files or in the RT Brachy Applications Setup (Table B6). These are set as follows:

| **Tags** | **Description** | **Example Value** |
|---|---|---|
| **SOP Common** | | |
| 0008,0005 | Specific Character Set | “ISO_IR 100” – Latin alphabet No. 1 |
| **General Study** | | |

| | | | |
|---|---|---|---|
| 0020,000d | | Study Instance UID | Study ID, see above |
| **RT Series** | | | |
| 0008,0060 | | Modality | “RTPLAN” |
| 0020,000e | | Series Instance UID | Series ID, see above |
| 0020,0011 | | Series Number | “1” |
| 0008,1070 | | Operator’s Name | “” |
| **RT General Plan** | | | |
| 300a,0002 | | RT Plan Label | “Testplan” |
| 300a,0006 | | RT Plan Date | "20190101" |
| 300a,0007 | | RT Plan Time | "000000" |
| 300a,000c | | RT Plan Geometry | Eclipse:<br>“PATIENT” (requiring linked structure set [300c,0060]).<br>Oncentra:<br>“TREATMENT_DEVICE” |
| 300c,0060 | | Referenced Structure Set Sequence | 0x0008,0x1150<br>0x0008,0x1155 |
| | 0008,1150 | Referenced SOP Class UID | "1.2.840.10008.5.1.4.1.1.481.3" - Structure Set Storage Class |
| | 0008,1155 | Referenced SOP Instance UID | Catheter UID – See Above |
| **RT Fraction Scheme** | | **Fraction Group Sequence (300a,0070)** | |
| 300a,0071 | | Fraction Group Number | 1 |
| 300a,0078 | | Number of Fractions Planned | 1 |
| 300a,0080 | | Number of Beams | 0 |
| 300a,00A0 | | Number of Brachy Application Setups | 1 |
| 300C,000A | | Referenced Brachy Application Setup Sequence | 300c,000c |
| 300C,000A | 300c,000c | Referenced Brachy Application Setup Number | 1 |

Table B6: Remaining tags.

**B.IV. Clinical Software Compatibility**
**B.IV.I. Catheter Reconstruction**

Both Oncentra TPS and Eclipse are dependent on catheter reconstruction information in order to import an RT Plan. These catheters are written as structures in RT Struct format, defined as a series of points. Eclipse requires that these structures be written in a standard RT Struct file, either alone or as part of the regular structure set. Oncentra TPS require the same information to be included as part of private sequence (300f,1000).

Control over the vtkGDCMPolyDataWriter class is limited, such that for the catheter reconstruction the file must be reopened to customize certain tags. First, in order to link the catheter reconstruction information to the plan, tag (0008,1155) Referenced SOP Instance UID ([ROI Contour -> (3006,0039) ROI Contour Sequence -> (3006,0040) Contour Sequence -> (3006,0016) Contour Image Sequence] is set to be equal to (0020,000e – Series Instance UID from the image DICOM). Also, tag (0008,0018 - SOP Common – SOP Instance UID) is set to a newly generated catheter ID, linked to RT General Plan: Referenced Structure Set Sequence: 0x0008,0x1155 – Referenced SOP. For Eclipse, this same UID is also written to tag (0002,0003 – SOP Instance UID) in the meta information. Next, the Contour Geometric Type (3006,0042) is switched from "CLOSED_PLANAR" to "OPEN_NONPLANAR". Finally, for incorporation into the RT Plan (see below), value representations within the dataset are recursively changed to explicit based on the public GDCM data dictionary.

**B.IV.II. Other tags**

In Oncentra TPS, the offset variable is calculated as Source applicator length (300a,0296) minus catheter path length. Source applicator length is therefore set equal to digitized catheter length to yield an offset of zero.

Oncentra requires private tags to be included that are not present in the DICOM specification (Table B7).

| **Tags** | | **Description** | **Example Value** |
|---|---|---|---|
| 3005,0010 | | Private Creator | "MDS NORDION CALCULATION" |
| 3007,0010 | | Private Creator | "MDS NORDION OTP EM" |
| 3007,1000 | | DAbsDoseGyAt100% | 5.0 |
| 300b,1000 | | Referenced RT Plan ROI Number | ROI number as written to structure set, indexed at 0 |
| 300f,0010 | | Private Creator | "NUCLETRON" |
| 3007,100b | | Normalization ROI info sequence | |
| 3007,100b | 3007,0010 | Private Creator | "MDS NORDION OTP EM" |
| 3007,100b | 3007,1015 | Normalization Distance | 0 |
| 3007,100b | 3007,1016 | Normalization Factor | 1 |
| 3007,100b | 3007,1018 | Normalization F-Factor | 0.37 |

Table B7: Other Oncentra required private tags.

Eclipse requires some tags to be included in the plan file that are not listed as mandatory in the DICOM specification (Table B8).

| **Tags** | **Description** | **Example Value** |
|---|---|---|
| 300a,0070 | Fraction Group Sequence and sub-tags (300a,0071), (300a,0078), (300a,0080), (300a,00a0) | 1, 1, 0 ,1 |
| 300a,0290 | Source Applicator Number | 1 (assuming one applicator) |
| 300a, 0218 | Active Source Diameter | "2" |
| 300a, 021a | Active Source Length | "5" |

Table B8: Eclipse mandatory tags listed as "Optional" in the DICOM specification.

## Appendix C: Technical Implementation Details

All the classes and functions mentioned below were implemented using VTK 8.2 (37) and Qt 5.12 (35).

### DICOM Data

Image files are read using a Vtk-GDCM data reader. They are sorted using GDCM IPPSorter (Image Position Patient sorter) which sorts according to the Image Plane Module of the DICOM standard, determining the ordering of the axial images and the z-spacing between them. VtkImageData objects are then generated, a topologically and geometrically regular data structure. VtkImageData only requires the spacing, dimensions and origin coordinates of the voxel grid, such that 3d coordinates for every other voxel do not have to be specified. The origin of this vtkImageData object is then set, again accessing the Image Plane Module. This image is then flipped along the y-axis using vtkImageFlip, because vtkDICOMReader rasterizes its output from bottom to top, while DICOM is rasterized top-to-bottom. This vtkImageData is put through a vtkLookupTable with vtkImageMapToColors, used to map image scalar values into RGBA for display. Saturation is set to zero (as is hue), to ensure that colors generated are all grey-scale, and alpha is set to 1, such that colors have full opacity. This data is then transformed to a vtkActor and added to vtkRenderers on each orthogonal plane for display. Each plane view window therefore has access to the entire image data actor, and the display extents are readjusted as the user scrolls through the image. DICOM RT files are read using the vtkGDCMPolyDataReader class, capable of reading DICOM RT files and outputting a series of vtkPolyData objects, one for each structure in the RT file. These PolyData objects consist of a point list and a series of polygons sorted by axial slice, each used to instantiate a DICOMStructure class which are stored in a vector in the DICOMStructureSet class.

### Basic Contouring

Contour segments are stored as planar polygons using vtkPolyData, specifically one vtkPolyLine with associated vtkPoints. Each of the three PlaneViewWidget subclasses maintains its own set of contour data, a vector with one entry per slice. Since structures sometimes branch into multiple segments on the same slice, each slice can store multiple PolyData segments. When multiple segments exist, the separate vtkPolyData representations are merged into one using the vtkAppendPolyData filter, such that ultimately only one vtkPolyData is saved per slice. Checks are performed when adding to or subtracting from the contour, to determine if any previously separated segments have been merged or any previously continuous segments split, respectively.

The contour polygons are generated differently depending on the contouring mode. When in freehand draw mode, polygons are generated by the CustomContourWidget class, inheriting from vktContourWidget, in which polygons are represented by the associated vtkOrientedGlyphContourRepresentation class. These classes, in conjunction with helper classes vtkPointPlacer and vtkContourLineInterpolator, allow the user to select nodes and draw lines between them. A vtkImageActorPointPlacer is used to constrain the user-selected points to the currently displayed plane of the patient image. These classes also enable contour morph functionality, clicking and dragging contour nodes. If the user clicks on an existing node, that node is used as the center point of the morph, otherwise a new node is created. When saving freehand contour data, polygons are pulled from the CustomContourWidget as vtkContourRepresentations, the converted to vtkPolyData.

Brush functionality, on the other hand, is implemented in the InteractorStyle class, inheriting from vtkInteractorStyleImage. 50-sided polygons are used to approximate circles, centered at the cursor location, and are boolean unioned to the active contour. Due to vtkBooleanOperationPolyDataFilter not supporting boolean operations on coplanar polygons, RapidBrachyMCTPS transforms vtkPolyData to QPolygonF data structures to make use of Qt's boolean operations, and then transform the data back to vtkPolyData for storage. Since boolean operations are continuously performed in brush mode, the data is always ready to be saved directly as vtkPolyData.

**GUI**

GUI elements are implemented with the Qt software development kit. Elements are arranged using the QLayout class and its children, which specify the relative positions of Qt Widgets. Various widgets are central to Qt functionality, and all inherit directly or indirectly from the QObject class. When a class inherits from QObject and contains the Q_OBJECT macro in the c++ header file, the Qt Meta-Object Compiler parses the class and generates additional code which adds functionality not provided by the standard c++ compilers, such as the signals and slots mechanism. Signals and slots are a powerful Qt feature, allowing for GUI elements to interact with each other without freezing the interface. One QObject-child class emits a signal, which is connected to one or more slots in the same or other QObjects, set to be processed in the next event loop. This event loop is also provided by Qt, and captures user keyboard and mouse events. See (34) for a more detailed discussion of Qt functionality.

**Contour Representations**

Each plane has its own vtkRenderer and InteractorStyle classes, and maintains its own set of planar polygon contours. A single ContourDrawer is responsible for saving these polygons, as well as converting to and storing the cumulative labelmap and closed surface representations. In order to speed up the conversions, polygons are converted to sectional labelmaps one slice at a time and added to a master labelmap representation in ContourDrawer. For slices from the axial plane, labelmaps are generated through basic linear extrusion. A vtkLinearExtrusion filter is applied, extruding all polygonal edges in the +z direction by exactly the z-spacing. The effect of this is to fill in the empty space between the axial slices with puck-like ribbons as illustrated in Figure C1.

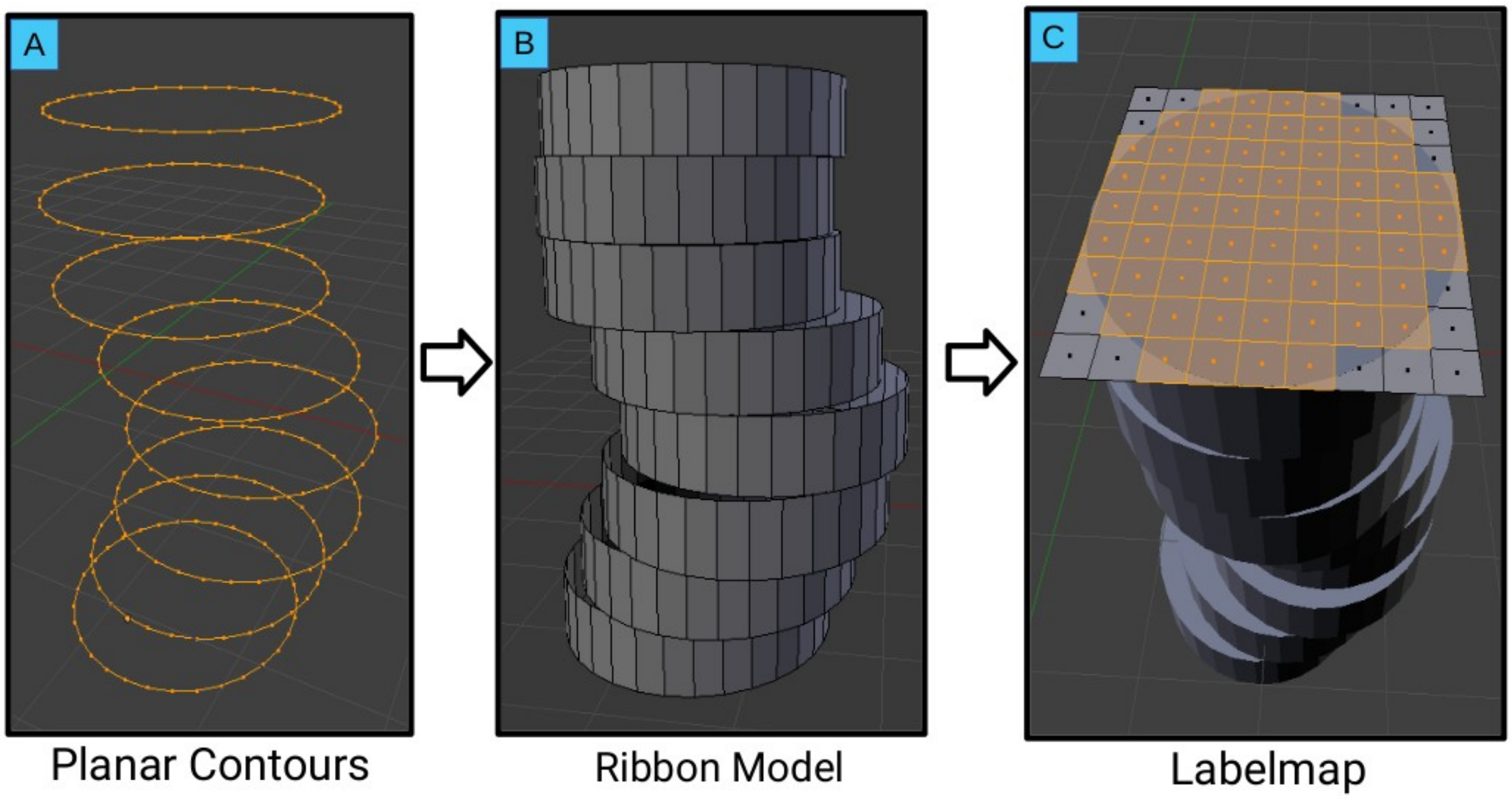

Figure C1: Conversion from planar contour to labelmap through linear extrusion to a ribbon model. A) planar contours B) ribbon model C) labelmap.

These pucks are also offset by half a z-spacing such that each slice lines up with the middle of the ribbon. A vtkPolyDataToImageStencil filter is then applied to the ribbon model, creating a binary image stencil. The tolerance for inclusion is set to zero, such that if a voxel's center does not fall within the surface mesh, it is not included in the binary mask. Next, a blank mask is created with the same geometry as the patient data (same origin, spacing and dimensions), and this mask is filled entirely with zeros. Finally, the blank mask is combined with the binary stencil using vtkImageStencil. The result is that any voxels included in the binary stencil are left as zeros in the blank mask, and any voxels not included are changed to a positive integer. If the user is adding to the structure, all voxels included in the sectional labelmap are included in the master labelmap. For slices in the sagittal and coronal planes, this conversion pipeline will not work, since the vtk filters used only support axial polydata. Sagittal and coronal contours must therefore be rotated into the axial plane. The stencil is then generated, and the stencil coordinates are flipped back into the original orientations. Again, any voxels included in these sectional labelmaps are added to the master structure labelmap.

If the user is subtracting from the structure, additional calculations must be performed in order to subtract the proper voxels and update the planar polygons on orthogonal axes. First, the user draws the subtraction polygon, detailing the area they wish to be removed from the structure. A hybrid representation is then created, a boolean union of the converted labelmap outlines with the original planar polygons. The polygonal difference between the subtraction polygon and the hybrid representation is then obtained. This difference is converted to a sectional labelmap as above, and any voxels included in this differential sectional labelmap are subtracted from the master labelmap. The outlines of these voxels are also obtained, and are subtracted from the planar polygons on the orthogonal axes. This is accomplished by iterating through the labelmap, and checking which voxel edges are at the boundary of inclusion and exclusion. These voxels edges are then reconstructed into a continuous loop delineating the mask boundary.

Finally, we create closed surfaces through Vtk's marching cubes algorithm, the three-dimension version of the marching squares algorithm, as presented in Figure C2. Marching squares proceed by considering each four adjacent voxels, and those of them that are inside or outside the labelmap. With binary inclusion, there are only 16 possible topological combinations. Lines are drawn based on comparison to a case list of these 16 possibilities. Marching cubes works analogously, but in three dimensions by comparing sets of 8 adjacent voxels with 256 possible cases.

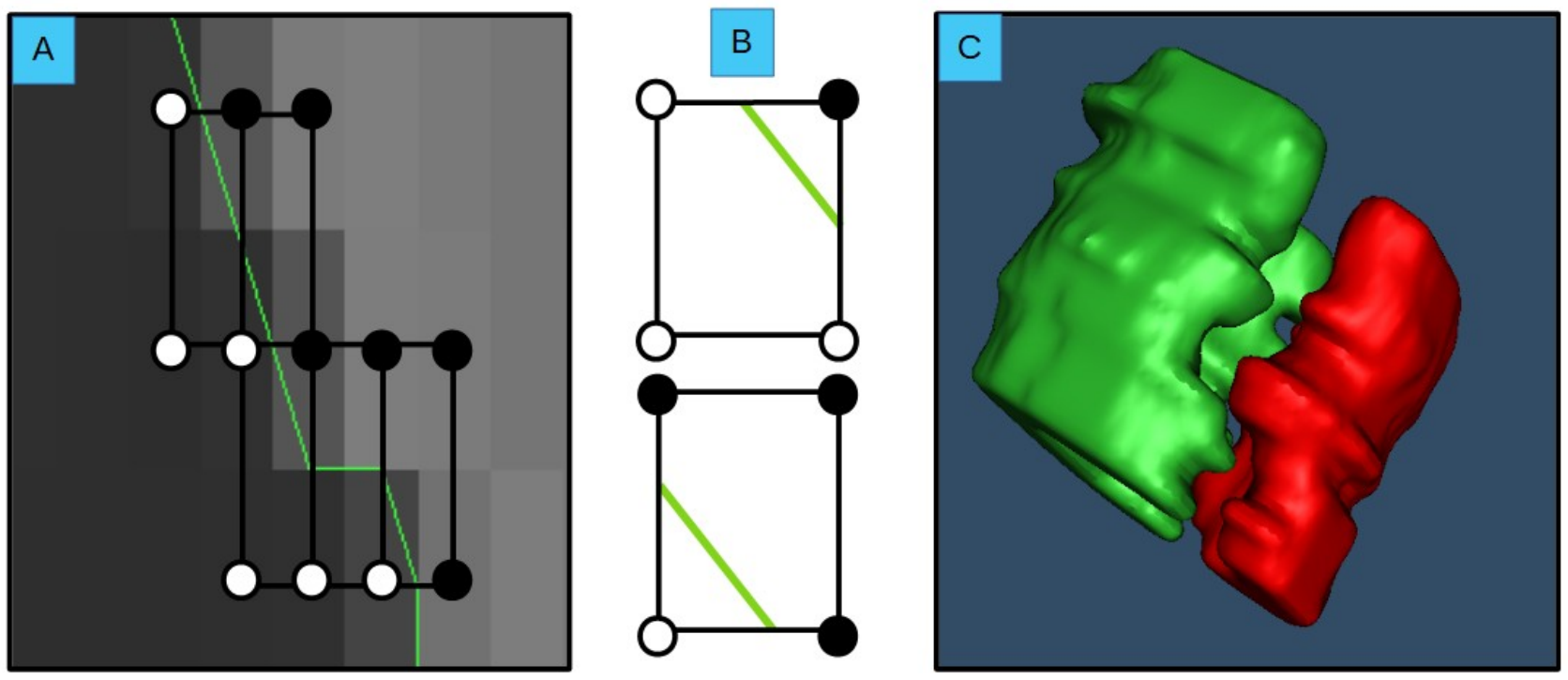

Figure C2: Conversion from labelmap to closed surface takes place through marching cubes algorithm, analogous to the 2D marching squares algorithm A) applied marching squares calculation B) marching squares two example cases C) closed surface representation.

Each plane widget class has its own vtkRenderer, and therefore renders its own contour actors. When the user panes through the image slices, all previously present actors are removed and new actors are calculated for each structure in separate threads. If polygonal data exists for a particular slice, the polygons are rendered as lines. If labelmap voxels are included on that slice, they are overlaid with a semi-transparent mask. If no polygonal data exists but labelmap voxels are still included, i.e. they have been added on an orthogonal plane, the marching squares algorithm is used to generate polygonal lines for display.

Polygonal data is converted from vtkPolyData to a vtkActor using a vtkPolyDataMapper. Labelmap mask actors are generated by passing the structure mask through a vtkLookupTable such that included voxels are semi-transparent and excluded voxels are fully transparent. The output of this lookup table is then passed through a vtkImageMapToColors filter, which is passed to a vtkImageActor for rendering. Finally, when generating new contours from labelmaps, vtkExtractVOI is used on the mask to return only the voxels from the currently displayed plane. The marching squares algorithm is then applied. Specifically, the extracted slice is passed through a vtkContourFilter, which generates line segments delineating the boundary between zero and background mask voxels. These line segments are passed through a vtkPolyDataMapper to create a vtkActor for rendering.

RapidBrachyMCTPS enables interpolation in all three axes via paired parametric tuples as shown in Figure 8. This method requires that planar polygonal data has been reordered such that points on one slice are listed in the same order as corresponding points on another slice. This reordering takes place every time a polygon is saved to the master representation. First, the polygon is scanned for the point with maximum summed world position coordinates (x + y + z). Next, the signed area of the contour is calculated, which is dependent on the orientation of the polygon (counterclockwise polygons give positive values, clockwise give negative). The points are then reordered, starting from the maximum summed point, and winding in a direction that ensures a positive signed area.

When a slice is selected for interpolation, nearest contours are found in positive and negative directions along the axis of interest. When multiple polygon segments appear on the same slice, it is necessary to calculate segment correspondence. Correspondence is established by comparing the average point coordinates of each polygonal segment. The best match between the coordinate averages is established on the first side of the interpolation. It is possible that multiple segments on the first side are matched to the same segment on the second side, leaving unmatched segments on the second side. These unmatched segments are then compared to and matched with segments on the first side, such that there is one-to-one correspondence between all segments. The software does not currently support slice interpolation between slices with differing numbers of contour segments.

Next, the matched segments are added point-by-point to a vtkTupleInterpolator, parameterized by cumulative contour length between 0 and 1. For instance, if a contour has points at 60% and 70% of the total contour length, those points will be passed in as tuples at parameter values 0.6 and 0.7, respectively. When interpolating, new points are sampled at regular intervals from the vtkTupleInterpolator, which linearly interpolates the adjacent points on either side of the passed-in parameter. For instance, for the contour described above, a parameter of 0.62 would return a tuple interpolated with 80% weight from the 0.6 point coordinates and 20% weight from the 0.7 point coordinates.

If a structure contains a keyhole, this entire pipeline occurs recursively. Keyholes are rearranged, correspondence between them is established, interpolation is performed, then finally the keyhole is reconnected to the outline. The software does not currently support slice interpolation between slices with differing numbers of keyholes.

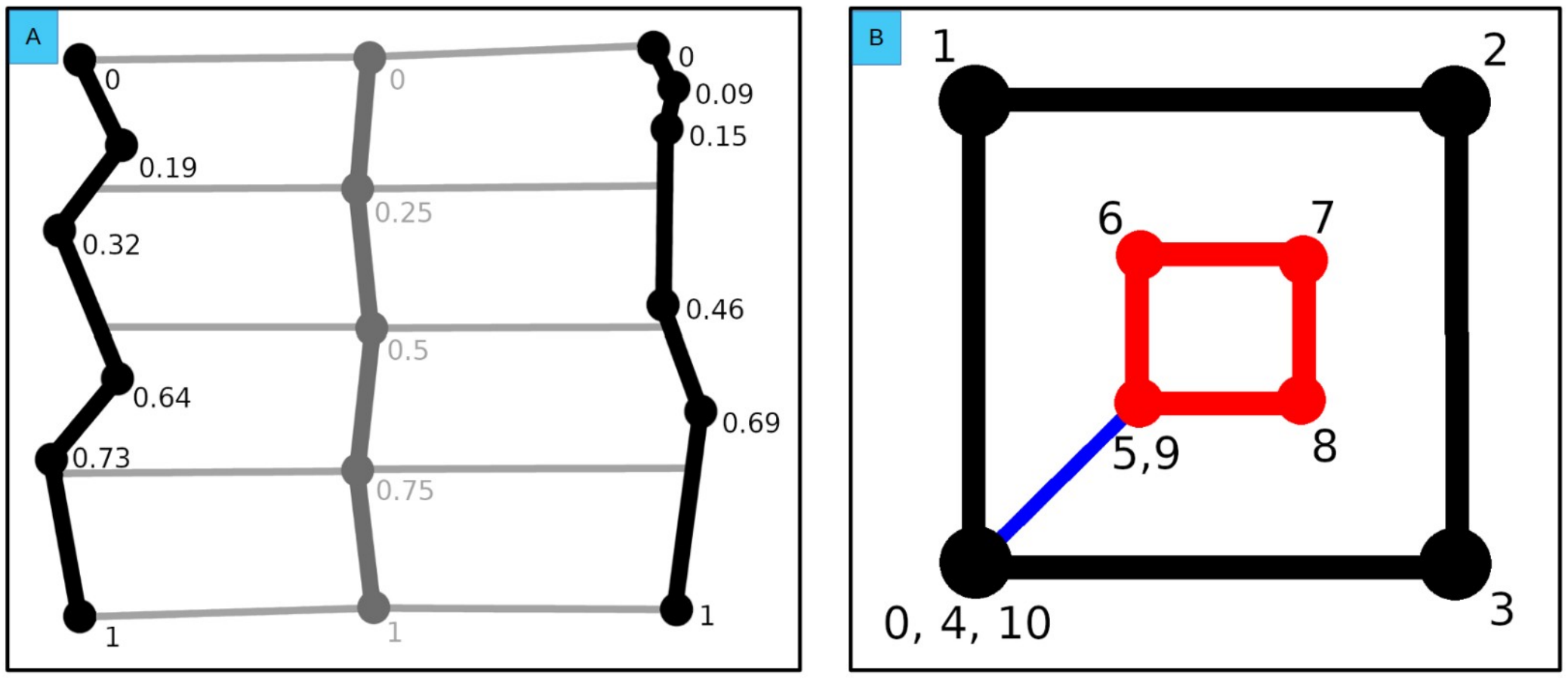

Figure C3: Linear interpolation via paired parametric tuples. A) unwrapped parameterized polygonal lines B) keyhole interpolation: black outline and red keyhole are split at blue degenerate line, interpolated separately, then rejoined.

## Appendix D:Applicator Geometries

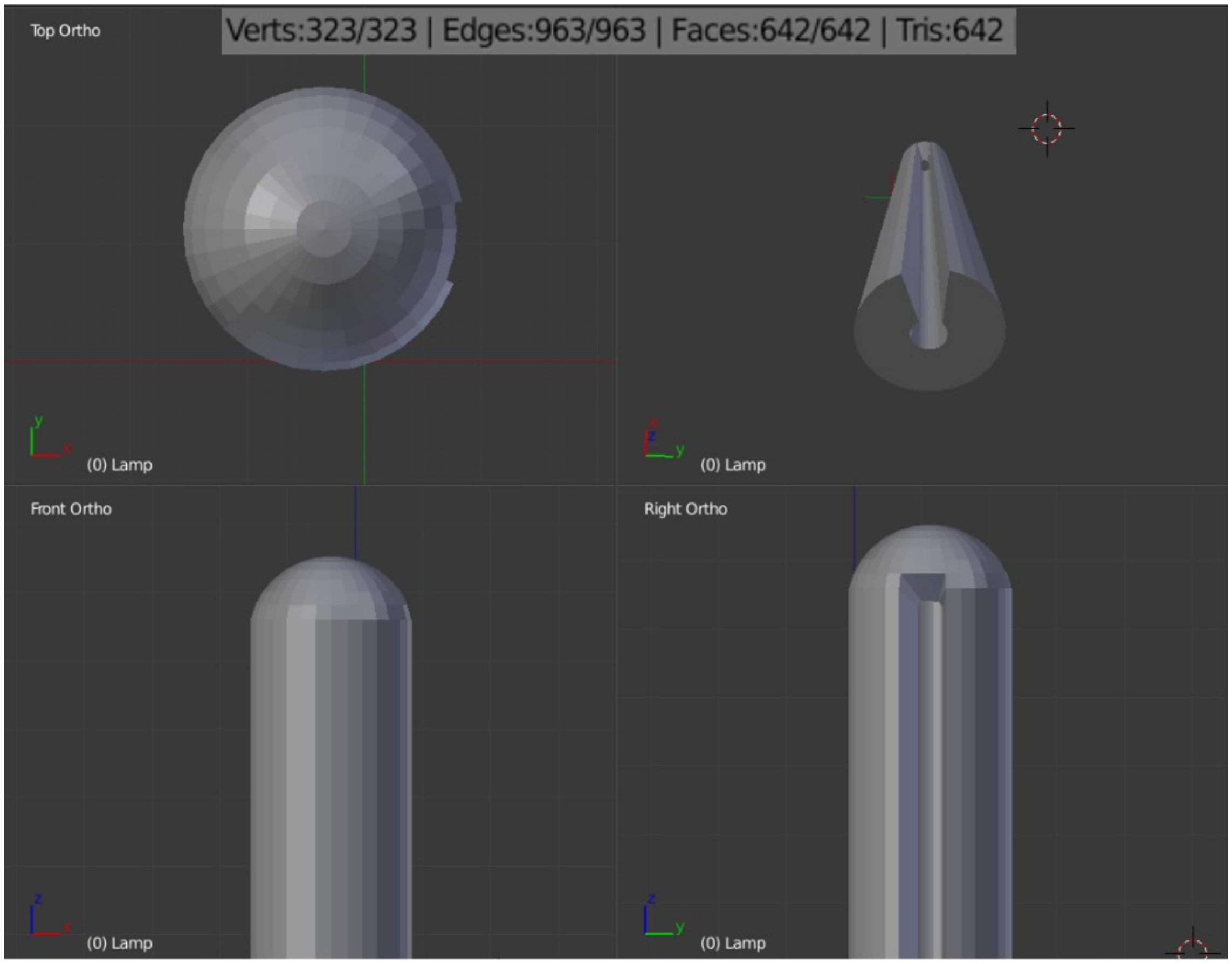

Figure D1: Grooved applicator similar to a shielded cervical brachytherapy applicator, modelled in Blender.

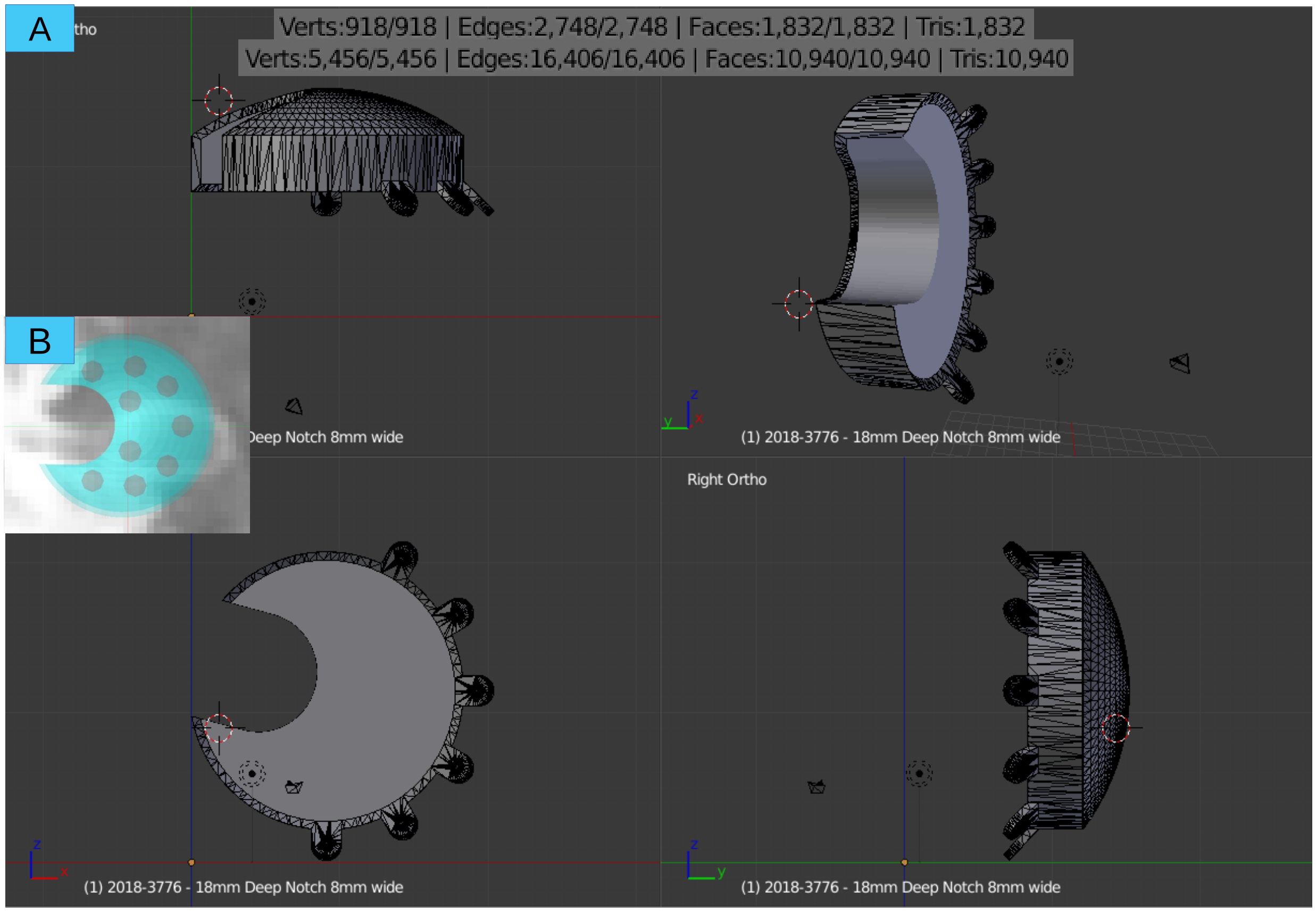

Figure D2: Eye plaque applicator. A) eye plaque provided by Trachsel Dental Studio, silastic insert filled in using Blender. B) dwell positions numerically placed at mid-line within the insert.

**Appendix E: Monte Carlo Parameters**

| **Property** | **Description** | **Reference** |
|---|---|---|
| **Software** | Geant4, RapidBrachyMCTPS. | (21),<br>(48),<br>(49) |
| **Validation** | Previously validated. | (21) |
| **Geometry** | Voxelized geometry (egsphant) extracted from DICOM CT images and DICOM RT Structure Set files. | |
| **Materials** | Heterogeneous (TG-186), with elemental composition of tissues and CT-to-density conversion as presented in table E2. | (8),<br>(1),<br>(50),<br>(51),<br>(6) |
| **Source Description** | microSelectron v2 or SelectSeed source geometries.<br>Explicit simulation of radioactive decay using photon decay spectra from ENSDF. Source positions and orientations imported from DICOM RT Plan files or defined in custom RapidBrachyMCTPS files. | (52),<br>(53),<br>(Elekta Brachytherapy, Veenendaal, The Netherlands) |
| **Cross Sections** | EPDL97, EEDL97, EADL97. | (54),<br>(55),<br>(56) |
| **Transport Parameters** | PENELOPE low-energy electromagnetic physics list with default transport parameters. Electron transport off. Production cut: 0.1 mm. | |
| **Variance Reduction Technique** | Track-length estimator using mass-energy absorption coefficient library provided in RapidBrachyMCTPS. | (21),<br>(57),<br>(6) |
| **Scored Quantities** | Absorbed dose (collisional kerma approximation) scored to water or medium. Voxel size same as CT voxel dimension. | |
| **Number of Histories/Statistical Uncertainty** | $10^7$ radioactive decays per dwell position. Type A uncertainties <1% for voxels inside 100% isodose lines. | |
| **Statistical Methods** | History-by-history method. | |
| **Postprocessing** | Dose to voxels converted into dose-volume parameters using RapidBrachyMCTPS. | (21) |

Table E1: Monte Carlo simulation methods based on the recommendations of TG-268. Modified with permission from Famulari *et al*.

| **Material** | **Density (g/cm$^3$)** | **Hounsfield Unit (HU)** |
|---|---|---|
| Air | 0 | < -1024 |
| Adipose | 0 - 0.98 | -1024 - -57 |
| Water | 0.98 - 1 | -57 - 0 |
| Soft Tissue | 1 - 1.052 | 0.0 - 14 |
| Rectum | 1.052 - 1.094 | 14 - 69 |
| Bladder | 1.094 - 1.155 | 69 - 217 |
| Prostate | 1.155 - 1.824 | 217 - 1230 |
| Bone | 1.824 - 3 | 1230 - 2000 |
| Metal | 3 - 19 | > 2000 |

Table E2: CT HU to density and material calibration curve used by McGill University Health Center Medical Physics Unit. Densities are linearly interpolated according to HU.

| **Material** | **Element (% mass)** | | | | | | | | | | | | |
|---|---|---|---|---|---|---|---|---|---|---|---|---|---|
| | **H** | **C** | **N** | **O** | **Na** | **Mg** | **P** | **S** | **Cl** | **Ar** | **K** | **Ca** | **Zn** |
| **Air** | 0.1 | | 75.0 | 23.6 | | | | | | 1.3 | | | |
| **Adipose** | 11.4 | 59.8 | 0.7 | 27.8 | 0.1 | | | 0.1 | 0.1 | | | | |
| **Water** | 11.2 | | | 88.8 | | | | | | | | | |
| **Soft Tissue** | 10.5 | 25.6 | 2.7 | 60.2 | 0.1 | | 0.2 | 0.3 | 0.2 | | 0.2 | | |
| **Rectum** | 6.30 | 12.08 | 2.21 | 78.81 | 0.11 | | 0.079 | 0.12 | 0.15 | | 0.11 | | |
| **Bladder** | 10.89 | | | 88.51 | 0.25 | 0.02 | 0.10 | | | | 0.21 | 0.02 | |
| **Prostate** | 9.76 | 9.11 | 2.47 | 78.1 | 0.21 | | 0.1 | | | | 0.2 | 0.023 | 0.008 |
| **Bone** | 3.4 | 15.5 | 4.2 | 43.5 | 0.1 | 0.2 | 10.3 | 0.3 | | | | 22.5 | |

Table E3: Elemental composition of tissue types. “Metal” not included in table due to non-overlapping elements. “Metal” composition is Fe 67.92%, Cr 19.00%, Ni 10.00%, Mn 2.00%, Si 1.00%, C 0.08%.

## Appendix F: Terminology

Brachytherapy: radiation oncology treatment modality where the radioactive source is placed in close proximity to the tumour.

Monte Carlo: stochastic simulation method, samples repeatedly from a probabilistic distribution in order to obtain a low-variance representation of that distribution.

Catheter: hollow cylindrical object in which radioactive sources are placed.

Dwell position: positions within a catheter where radioactive sources can rest.

Dwell time: the amount of time that a radioactive source is left in a given dwell position.

Target: the tumour or other structure targeted by the treatment plan in order to deliver a high dose. A clinical target volume (CTV) is often defined, which accounts for tumour margins that may not appear on the image, as is a planning target volume (PTV), which expands the volume further to allow for uncertainties in treatment planning.

Organs at risk: an organ or other structure which should receive a low dose to avoid toxicities.

Vector: used in this paper to mean std::vector from the c++ standard library.